\documentclass[onecolumn]{emulateapj}

\usepackage{array,graphicx,natbib}

\renewcommand{\vec}[1]{\mathbf{#1}}

\begin{document}

\title{Protoplanet Dynamics in a Shear-Dominated Disk}

\author{Benjamin F. Collins and Re'em Sari}
\affil{California Institute of Technology, MC 130-33,
	Pasadena, CA 91125}
\email{bfc@tapir.caltech.edu}

\begin{abstract}
The velocity dispersion, or eccentricity distribution, of protoplanets
interacting with planetesimals 
is set by a balance between dynamical friction and viscous stirring.
We calculate analytically the eccentricity distribution function of
protoplanets embedded in a cold, shear-dominated planetesimal swarm.
We find a distinctly non-Rayleigh distribution with a simple 
analytical form.  The peak of the
distribution lies much lower than the root-mean-squared value,
indicating that while most of the bodies have similarly small
eccentricities, a small subset of the population contains most of the
thermal energy.  We also measure the shear-dominated
eccentricity distribution using numerical simulations. 
The numerical code treats each protoplanet explicitly and adds
an additional force term to each body 
to represent the dynamical friction of the 
planetesimals.  Without fitting any
parameters, the eccentricity distribution of protoplanets in the N-body
simulation agrees with the analytical results.
This distribution function provides a useful tool
for testing hybrid numerical simulations of late-stage planet formation.
\end{abstract}


\section{INTRODUCTION}

Terrestrial planets, ice giants, and the cores of the gas giants are
thought to form by accretion of planetesimals into protoplanets.  
The protoplanets emerge from the swarm of planetesimals after
an epoch of runaway accretion.  The subsequent dynamics of the 
protoplanets set several important features of the final
planetary configuration, such as the mass and number of planets
or cores.  It is difficult to constrain this evolution, however, 
without constraining the properties of the disk in which
they are embedded.

One important yet uncertain parameter 
is the size of the planetesimals, the building blocks.
The outer Solar System and the later stages of formation in the 
inner Solar System likely lack gas, allowing the formation of
kilometer-size bodies through gravitational instabilities.
Those bodies collide and grind each other down to even smaller sizes 
in a collisional cascade. The existence of bodies small enough to damp 
their own velocity dispersion is an inevitable conclusion from the
existence of Uranus and Neptune \citep{GLS04}.  Without such small bodies, 
the ability of a growing protoplanet to gravitationally focus 
the planetesimals becomes inefficient, and the growth
timescale becomes too long, of order $10^{12}$~years in the outer solar
system. 

The unavoidable influence of the planetesimals make numerical
studies of planet formation difficult to carry out accurately.
Despite modern computational power, an integration of  
the equations of motion for each body in a 
protoplanet and planetesimal swarm is impossible. 
Even without allowing planetesimal fragmentation, 
the number of kilometer-size bodies needed to comprise a 
Neptune size mass is humongous, of order $10^{12}$. 
\cite{KI96} performed numerically feasible but physically less appropriate 
N-body simulations of a proto-planetary disk 
in which the size of planetesimals
is larger than the value required to form the ice giants of our 
Solar System.  Although interesting from a dynamical viewpoint,
the results of such simulations can not be extrapolated to
the scenario of smaller planetesimals since they
lack collisional damping.

An alternative numerical approach to studying these systems 
is a coagulation code~\citep{Lee00,KL98} in which the
bodies are divided into size bins and the interaction 
of each pair of bins is calculated statistically. 
This approach fails once the number of bodies in any bin
is not sufficiently large.
\cite{KB05} have developed a hybrid code that treats
planetesimals statistically while a small number of large bodies are
integrated individually. 

In this paper, we examine the processes that shape the eccentricity 
distribution of the large bodies. We assume, simply,
that the planetesimals constitute a cold disk due to sufficiently frequent
collisions. As a first step, we include the dynamical 
friction that the planetesimals exert on the large bodies but
ignore the much slower process of their accretion onto 
those bodies.  The rates of cooling from dynamical friction and 
heating from mutual excitations are discussed in \S \ref{roughrates}.
We write a Boltzmann equation to
show the change in the distribution function of eccentricities due to
each process in \S \ref{analytic}, and discuss the solution
to that equation.  In \S \ref{nbody} we
present the results of complementary N-body simulations designed to
measure the eccentricity distribution directly.  A discussion of the
results follows in \S \ref{conclusions}.

\section{SHEAR-DOMINATED COOLING AND HEATING RATES}
\label{roughrates}

The eccentricities of the protoplanets represent
a kind of ``thermal'' energy in their orbits, relative to perfectly circular 
motion.  The extra non-circular velocity itself varies in magnitude 
and direction over an orbital period; it is simpler to 
use the eccentricity, a constant of motion for the two-body problem.  
Specifically, we  calculate the vector eccentricity,

\begin{equation}
\label{velvector}
\vec e = \frac{\vec v \vec{\times} \vec H}{G M_p} - \frac{\vec{r}}{r}.
\end{equation}

\noindent
This expression relates the eccentricity of the particle, $\vec e$, 
to the particle's position, $\vec r$, its velocity, $\vec v$, its 
orbital angular momentum vector, $\vec H$, and its mass, $M_p$.
%This vector always lies in the same plane as $\vec v$ and $\vec r$,
%and its magnitude is simply the orbital eccentricity.
In general,  a protoplanet can have an inclination relative to the disk plane,
and the eccentricity vector can have three components.  However,
we show in \S \ref{inclinations} that the shear-dominated 
regime strongly inhibits the growth of inclinations.
Two dimensions then suffice to describe the configuration space
of $\vec e$.

We use the quantity of the Hill radius repeatedly in this work; for reference
we define its value as 

\begin{equation}
R_H~ \equiv~ 
\left(\frac{M_p}{3 M_{\odot}}\right)^{1/3} a ~ =~ R/\alpha,
\end{equation}

\noindent
where $M_p$ is the mass of a particle,  $a$ is its semi-major axis, 
$R$ is its radius, and

\begin{equation}
\alpha ~ = ~\left(\frac{M_p}{3 M_{\odot}}\right)^{1/3} \frac{R}{a} ~ = ~
\left(\frac{\rho_p}{3 \rho_{\odot}}\right)^{1/3} \frac{R_{\odot}}{a}.
\end{equation}

\noindent 
The Hill radius in turn specifies an eccentricity, the Hill eccentricity,

\begin{equation}
e_H = R_H / a.
\end{equation}

\noindent
We restrict this study to disks where the majority of the 
bodies have eccentricities lower than $e_H$,
known as the shear-dominated regime.

For most of this paper, we employ the ``two groups''
approximation~\citep{WS89,GLS04} and split the disk into 
two uniform populations.  One group is the numerous smaller bodies,
or ``planetesimals.''  We denote their
surface mass density as $\sigma$.
%, and the mass and 
%radius of an individual 
%planetesimal as $m$ and $s$.
The other group, the ``protoplanets'', consists of 
the bodies that dominate the 
excitations of the disk particles.  Each protoplanet has a radius $R$, 
mass $M$, and eccentricity $e$.  We write the total 
surface mass density in protoplanets
as $\Sigma$.  We assume 
$\sigma > \Sigma$ although this is not always true in the final stages of 
planet formation \citep{GLSfinalstages}.

\subsection{Eccentricity Excitation of Protoplanets}
\label{exciterate}

We analyze the interaction of two protoplanets
from a frame rotating with a reference orbit
at a semi-major axis $a$.  
The difference between the Keplerian angular velocity at each radius
induces a shearing motion between particles on nearby circular orbits.  
For an orbit interior to $a$ by a distance $b$,

\begin{equation}
\label{relfreq}
\Omega_{\rm rel}(b) = \Omega(a+b) -\Omega(a) 
\approx  \frac{3}{2} \Omega \frac{b}{a},
\end{equation}

\noindent
in the limit of $b \ll a$.  This angular frequency 
also specifies the rate of conjunctions for the two bodies 
with orbits separated by $b$.

The change in their eccentricity from each conjunction
can be calculated analytically for two nearly circular orbits
when $b \gg R_H$:

\begin{eqnarray}
\label{kickdist}
e_k& = &A_k~ e_H \left(\frac{b}{R_H}\right)^{-2}, \\
A_k & = & \frac{16}{3} \left[K_0(\frac{2}{3})+
\frac{1}{2}K_1(\frac{2}{3})\right] \approx 6.7.
\end{eqnarray}

\noindent
$K_0$ and $K_1$ are modified
Bessel functions of the second kind~\citep{GT80,PH86,DQT89}.
We note that 
$e_k$ refers to the perturbed body, while $e_H$ and $R_H$ 
refer to the Hill parameters of the perturbing one. 
The kick, viewed as a change in the eccentricity vector, is 
independent of the original eccentricity of the particle.
Its orientation is perpendicular to the line connecting the two 
protoplanets and the sun at conjunction; we therefore assume it
is random.

The eccentricity kick given by one
protoplanet is strongest for the particles 
that approach with an impact parameter on the order of the Hill radius.  
Interactions from a greater distance, however, occur more often.
In a shear-dominated disk, eccentricities
are small, $e \ll e_H$; to change that eccentricity significantly
only requires small perturbations.  These 
frequent but weaker perturbations dictate the overall velocity evolution of 
the protoplanets~\citep{GLS04,Raf04}. 

Specifically, the average differential rate that one protoplanet receives 
eccentricity kicks of strength $e$ from other protoplanets is given by

\begin{equation}
\label{diffkickrate}
d {\cal R}_{\rm ex}(e) = 2~ n_{\rm big}~ \frac{3}{2}
\Omega b(e)~\frac{d b}{d e}~ d e,
\end{equation}

\noindent
where $n_{\rm big}$ is the number surface density of protoplanets  
and $\frac{3}{2} \Omega b(e)$ is the velocity of encounters at those 
separations (given by eq. \ref{relfreq}).
The factor of two accounts for the 
combination of interior and exterior encounters.
The excitation rate of a protoplanet with eccentricity $e$
is then the rate of kicks comparable in magnitude 
to its current eccentricity:

\begin{equation}
\label{kickrate}
\left . \frac{1}{e} \frac{d e}{d t}\right|_{\rm ex} \sim e \left|\frac{d {\cal R}_{\rm ex}(e)}{d e}\right| 
\sim \frac{\Sigma \Omega}{\rho R} \frac{1}{\alpha^{2}} \frac{e_H}{e}.
\end{equation}

\noindent 
The inverse of this rate can be  
interpreted as the timescale for a protoplanet's eccentricity to change 
by an amount $e$.

\subsection{Dynamical Friction}
\label{dynfric}
As each protoplanet moves through the disk, it scatters and excites 
the eccentricities of the 
planetesimals that surround it.  Cold planetesimals that approach a
protoplanet with impact parameters of about a Hill sphere 
leave with $\sim m~ e_H$ of  
additional momentum.  This can either add to or subtract
from the eccentricity of the protoplanet depending on the 
relative orientation between the pre-encounter 
eccentricities of the protoplanet and planetesimal.
We write the net effect~\citep{GLS04}

\begin{equation}
M \frac{de}{dt}  \sim 
-n_s R_H e_H m(e_H+e) + n_s R_H e_H m(e_H-e),
\end{equation}

\noindent
for a number surface density of planetesimals $n_s$.
This formula yields the damping rate, or the inverse damping time,

\begin{equation}
\label{dynfriceq}
\tau_d^{-1} \equiv \left . \frac{1}{e} \frac{de}{dt}\right|_{\rm d.f.} =
%-
 C_d\frac{\sigma \Omega}{\rho R} \frac{1}{\alpha^2}.
\end{equation}

\noindent
Calculating the coefficient $C_d$ requires a more precise analysis of
planetesimal scattering.  We adopt the value $C_d \approx 10$
found by~\citet{Oht02}, who measure the coefficient
numerically.

\subsection{Planetesimal Interactions}

The distribution of planetesimal eccentricities does not
affect our results, as neither the excitation nor the damping rates
depend on their eccentricity as long as the 
planetesimals remain in a shear-dominated state.
In this work,
we focus on a range of parameters such that collisional cooling keeps
the eccentricities of planetesimals 
below $e_H$ and enforces the condition of shear domination.

\subsection{Inclinations}
\label{inclinations}
An orbit with a small inclination angle $i$ carries its particle
out of the disk plane on vertical excursions of size $\sim i a$.
An interaction that excites that particle's eccentricity also
affects its inclination, but with a magnitude inhibited
by the geometry of the distant encounter:

\begin{equation}
\frac{i_{k}}{e_k}  \sim  \frac{a~i}{b} ~ \Rightarrow~ \frac{i_k}{i} \sim 
\left(\frac{e}{e_H}\right)^{3/2}\frac{e_k}{e},
\end{equation}

\noindent
where $b$ is the impact parameter of the perturber and 
$i_k$ is the resulting change in inclination from an encounter.
In contrast, planetesimals just entering the Hill sphere of
a protoplanet damp the protoplanet's non-circular velocity in all dimensions;
no equivalent geometric factor inhibits the damping of inclinations.  
With the growth of inclinations suppressed, shear-dominated protoplanet 
disks are effectively two-dimensional~\citep{WS93,GLS04,Raf03c}.

\subsection{The Eccentricity Distribution - a Qualitative Discussion}
\label{qualitative}

The dynamical friction rate sets a characteristic time 
over which the eccentricities of all of the bodies are
changed significantly.  In this sense, the 
eccentricity distribution of the proto-planetary swarm 
is reset every $\tau_d$.
The excitation rate, however, varies with $e$.
Equating the excitation rate, equation \ref{kickrate}, and the damping rate,
equation \ref{dynfriceq}, yields an important reference value, $e_{\rm eq}$:

\begin{equation}
\label{eeq}
e_{\rm eq} \sim  \frac{\Sigma}{\sigma} e_H.
\end{equation}

\noindent
Statistically, each protoplanet receives one kick of this magnitude 
every damping timescale.

We deduce the distribution of eccentricities on each side of $e_{\rm eq}$ 
by examining the dependence of the kicking rate on eccentricity.
Excitation to $e \gg e_{\rm eq}$ requires a kick $e_k \gg e_{\rm eq}$.
Such strong kicks occur less often in one damping timescale 
than kicks of strength $e_{\rm eq}$ by a factor of $e_{\rm eq}/e_k$.
With fewer kicks to populate the high eccentricity distribution, 
the number of bodies with such eccentricities echoes the rate of kicks 
and falls off with eccentricity as $e^{-1}$\citep{GLS04}.

Kicks of order $e \ll e_{\rm eq}$ occur frequently in each damping
timescale, thereby overwhelming the effects of dynamical friction
on the lowest eccentricity bodies.
A dynamic equilibrium dominated by only the stirring mechanism
implies that kicks to and from
every eccentricity vector occur at the same rate.  For this 
to be true the distribution must be constant over
the configuration space.  The number of bodies with an eccentricity 
of order $e \ll e_{\rm eq}$ then scales
as the area of configuration space available, $\sim e^2$.

\section{A BOLTZMANN EQUATION}
\label{analytic}
In the following section we develop a differential equation to describe
analytically the distribution function of protoplanet eccentricities.
We construct this equation in the spirit of the Boltzmann equation,
examining the change in the number of bodies with a particular eccentricity
due to the effects of dynamical friction and viscous stirring.  

The space of possible eccentricities
is inherently two-dimensional (eq. \ref{velvector}),
since inclinations can be neglected (\S \ref{inclinations}).
Additionally, the interaction rates depend only on the
magnitude of the protoplanet eccentricity, forcing the distribution
function to share this dependence: $f(\vec e) = f(e)$.
The two-dimensional $f(e)$ is related to 
the number of bodies with velocity on the order of $e$ by its integral,
roughly $e^2 f(e)$.

Dynamical friction lowers the eccentricities of all bodies
proportionally to their eccentricity.  Equivalently, 
the number of bodies with a certain $e$ changes as the
protoplanets with that value are
damped to lower eccentricities and replaced by bodies from a
higher eccentricity.  We write this as:

\begin{equation}
\label{damprate}
\left . \frac{\partial f(e)}{\partial t}\right |_{\rm d.f.}  =  - ~ {\rm div} 
 \left( f(e) \frac{\partial \vec e}{\partial t} \right) 
  =  \frac{\partial f(e)}{\partial e} \frac{e}{\tau_d}
 + \frac{2 f(e)}{\tau_d},
\end{equation}
where we have used 
$\partial \vec e / \partial t = - \vec e /\tau_d$ for the effects
of dynamical friction.  

%As described in \S \ref{exciterate}, 
%an encounter between protoplanets gives each body a 
%``kick'' through eccentricity space.
At a given $\vec e$, particles are kicked
to a new eccentricity $\vec e_n$ at an average rate that
depends on the magnitude of the kick, $|\vec e_n - \vec e|$.  Also, particles
with an initial eccentricity $\vec e_n$ are kicked to $\vec e$
at the same rate.  The total flux of particles to and from a given
eccentricity is:

\begin{equation}
\left . \frac{\partial f(e)}{\partial t}\right |_{\rm kicks}
= 
 \int{\int{p(|\vec e_n -\vec e|) [f(e_n)- f(e)]~d^2 \vec e_n}}, 
\end{equation}

\noindent
where $p(e)$ describes the rate at which bodies 
experience changes in their eccentricities by an amount $e$.
This is the two-dimensional analog of the excitation rate
equation \ref{kickrate}:

\begin{equation}
\label{prate}
p(e) 
 =  \frac{1}{2 \pi~e} 
\left|\frac{\partial {\cal R}_{\rm ex}}{\partial e}\right| 
= A_k ~\frac{9}{16 \pi^2}  \frac{\Sigma \Omega}{\rho R} \frac{1}{\alpha^{2}}~ 
e_H~ \frac{1}{e^3}.
\end{equation}

The sum of the dynamical friction terms and the kicking integral
describes the dynamics of shear-dominated protoplanets
interacting with each other in a smooth disk of planetesimals.
The combined influence of these two processes can bring the 
protoplanets into an equilibrium state, where the number of particles with  
eccentricity $\vec e$ remains constant in time:

\begin{equation}
\label{fulleq}
 0 = 
\frac{\partial f(e)}{\partial e} \frac{e}{\tau_d} + \frac{2 f(e)}{\tau_d}
+ \int{\int{ p(|\vec{e}_n-\vec e|)~[ f(e_n)-f(e)]~d^2 \vec{e}_n}}.
\end{equation}

\subsection{The Solution}

We show in Appendix A that

\begin{eqnarray}
\label{thesolution}
f(e) &=& 
\frac{1}{2 \pi e_*^2} \left[1+\left(\frac{e}{e_*}\right)^2\right]^{-3/2}, \\
e_* &=& \frac{9 A_k}{8 \pi C_d} e_{\rm eq} 
\approx 0.24 \frac{\Sigma}{\sigma} e_H \nonumber
\end{eqnarray}

\noindent
satisfies the equilibrium
equation, equation \ref{fulleq} for all $e$.
This function is the equilibrium eccentricity distribution
of shear-dominated protoplanets.

The solid line in Fig.\ 1 shows a distribution function
for $\Sigma \approx 0.002 {\rm~ g~ cm^{-2}}$
and $\sigma = 0.1 {\rm ~ g~ cm^{-2}}$.  
Although the function formally extends above $e_H$, we stress
that it is only accurate for eccentricities $e \ll e_H$.  Both
the dynamical friction and the excitation rates 
(eqs.\ \ref{damprate} \& \ref{prate}) are not valid for $e \gtrsim e_H$.

Several moments of the distribution can be calculated
in terms of the only free parameters:

\begin{equation}
\frac{e_{\rm median}}{e_H} = 0.41 \frac{\Sigma}{\sigma},\quad 
\frac{\langle e \rangle}{e_H} = 0.24 \frac{\Sigma}{\sigma} \log (3 \frac{\sigma}{\Sigma 
}),\quad
\frac{\langle 1/e \rangle^{-1}}{e_H} = 0.24 \frac{\Sigma}{\sigma}.
\end{equation}

\noindent
According to equation \ref{thesolution}, 
$\langle e \rangle$ is infinite.  However, the 
largest single kick in eccentricity from an almost circular
protoplanet encounter is of order $e_H$.  Truncating the integral
at $e_H$ produces the logarithmic term in the expression
above.  Moments higher than the mean also diverge; realistically, they
are dominated by the bodies with the highest eccentricities, of 
order $e_H$.

It is easy to see that this solution, in the high- and low-eccentricity
limits, produces the same power-laws discussed in \S \ref{qualitative}.
In fact, it can be shown directly 
from equation \ref{fulleq} that any solution to
this differential equation reduces to those limiting power-laws.

\section{NUMERICAL SIMULATIONS}
\label{nbody}

Here we describe a direct measurement of the eccentricity distribution from
gravitational N-body simulations that include an additional force
to represent dynamical friction.

The N-body part of our simulation uses Gauss's
equations to evolve a set of orbital constants chosen to vary slowly under 
small perturbations.
A modified version of Kepler's equation produces the orbital phase for each
body at each time step.  The IDA solver from the SUNDIALS software
package~\citep{SUNDIALS} integrates the dynamical equations.
During close encounters of two protoplanets, we integrate
their motion relative to the center of mass of the pair.

We represent the planetesimal population in these simulations
with an extra force term that 
damps the non-circular velocities of the protoplanets at the rate
${\cal R}_d(e)$, given by equation \ref{dynfriceq}.  
An ad-hoc transition between the damping rate
for $e<e_H$ and the appropriate rate for $e > e_H$ 
prevents unphysical enhancements
of the damping force during close encounters.
The growth of protoplanets in mass 
due to planetesimal accretion is not included; the accretion rate is 
always lower than the dynamical friction rate and will not affect 
the eccentricity evolution~\citep{GLS04}.

Each simulation begins with the protoplanets on circular orbits
with random phases and random semi-major axes, within a
chosen annulus.  The average spacing between bodies, $M/(\Sigma 2 \pi a)$,
is several Hill radii.  
The protoplanets interact for several damping timescales $\tau_d$  before the 
distribution reaches equilibrium.  

We record the eccentricity 
of the protoplanets every $\Delta t \approx 0.1 \tau_d$
starting at about 100 $\tau_d$.  Several hundred 
orbits produces a well-populated histogram of eccentricities.
The bodies in the inner and outer edges of the 
disk are not measured, to avoid artificial boundary effects that 
inhibit excitations.  We bin the resulting eccentricities logarithmically.
Errors are assigned to each bin according to a Poisson distribution 
with the sample size defined as the product of the number 
of bodies measured and the sampling time in units of the 
damping timescale $\tau_d$.   Since each protoplanet suffers a
significant change in eccentricity every $\tau_d$,
one measurement of the eccentricity distribution  is 
independent from a previous measurement if they are 
separated in time by $\tau_d$.  We sample faster than $\tau_d$ 
to increase the resolution of the histogram slightly.

The statistical error bars do not take into account
the inhomogeneity of the protoplanet disk on small length scales.
Given a surface density, the mass of a single protoplanet
sets a typical radial separation between bodies.
This length scale corresponds to an eccentricity scale through
equation \ref{kickdist}  (in the simulations presented here, this value
is slightly below $e_H$).  
As the disk evolves, the viscous stirring causes migrations in
the semi-major axes of the particles that smooth the average
radial distribution.
If measured only over intervals shorter than the migration timescale,
the eccentricity distribution may vary for eccentricities above
the eccentricity set by the typical separation.  Fluctuations
from this effect are visible in Figs.\ 1 and 2.

Several simulations of disks with different protoplanet mass distributions
are presented below.

\subsection{Equal Mass Protoplanets}
\label{equalmass}

Figure 1 shows the eccentricity distribution measured from
a simulated disk of 120
equal mass protoplanets ($M = 2.5 \times 10^{-9} M_{\odot}$)
with surface densities $\Sigma \approx 0.002 {\rm ~g~cm^{-2}}$ and
$\sigma = 0.1 {\rm~g~cm^{-2}}$.  
A single population of protoplanets best 
reflects the ``two groups'' approximation we use
to derive equations \ref{kickrate} and \ref{dynfriceq}.
The analytic solution, equation \ref{thesolution},
for the same parameters in the simulation is superposed on Figure 1.  
While the overall match is not perfect, the shape of each curve is 
strikingly similar.   The two curves match extremely well if one is 
shifted by around 15 percent in the $e$ direction.  
This difference is attributable to the difficulty of
assigning a correct value of $\Sigma$ to the simulation
given a finite number of protoplanets. 

We note that there
are no free parameters in this comparison.  The numerical distribution
is a direct counting of the number of bodies within each eccentricity bin,
while a choice of $\Sigma$, $\sigma$, and $M$ completely specifies
the analytical curve.  

\begin{figure}

  \begin{center}

    \includegraphics[angle = -90, width = 0.7\columnwidth]{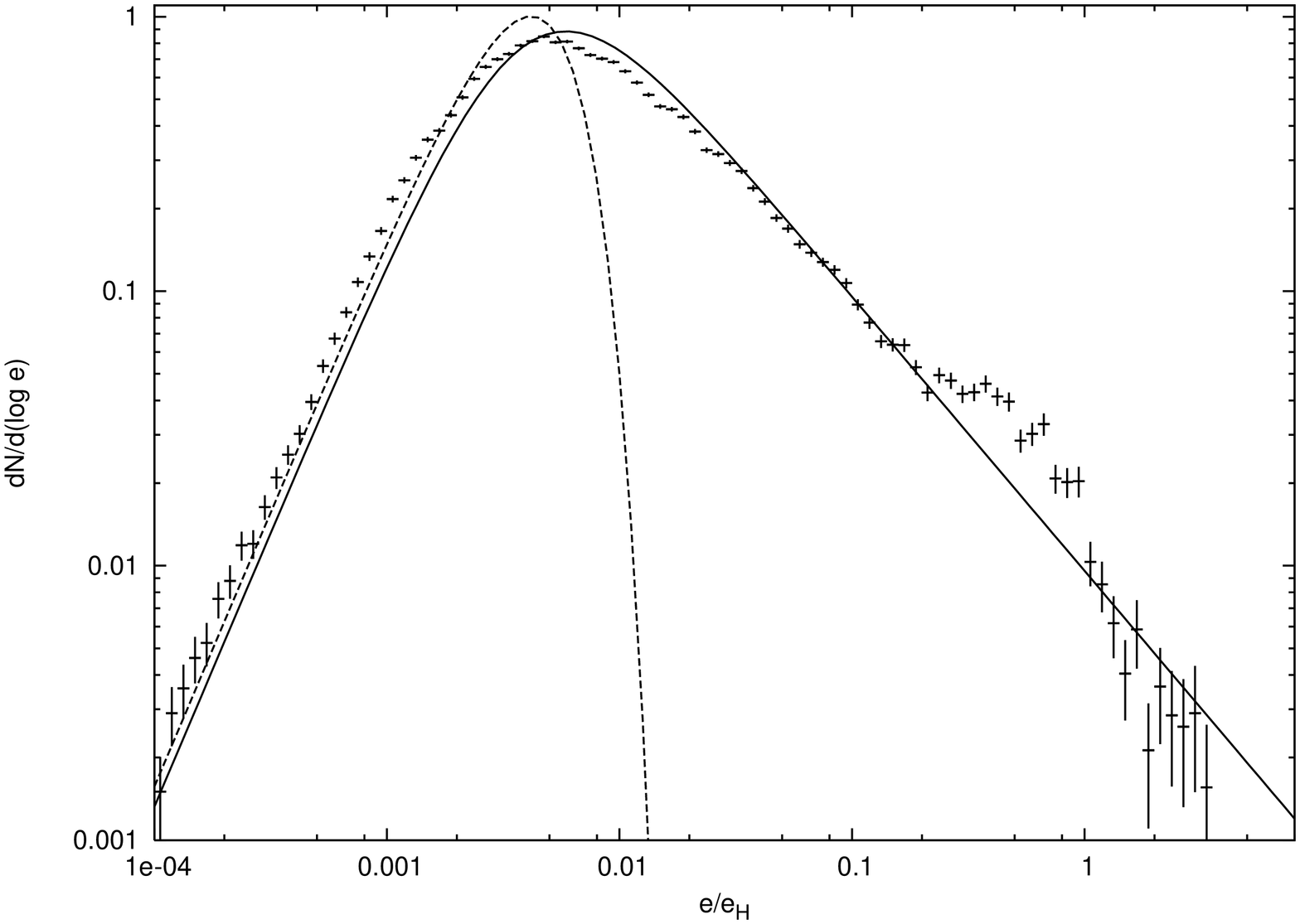}

  \end{center}
  
{\bf FIGURE 1:}  A plot of equation \ref{thesolution} superposed with
the results of a numerical simulation.  The simulated disk contains 
120 bodies of mass $M= 5 \times 10^{24} {\rm g}$, or 
$\Sigma \approx 0.002 ~{\rm g~cm^{-2}}$. A 
planetesimal surface density of $\sigma = 0.1 ~{\rm g~cm^{-2}}$
is included.
We assume each 
bin obeys Poisson statistics and assign errors based on a population size of
$N_b  N_{\tau_d}$, where $N_b$ is the number 
of bodies in the simulation, and 
$N_{\tau_d}$ is the duration of the simulation in units of damping timescales.
The solid line shows the distribution as given by equation \ref{thesolution}, 
using the same values of $\Sigma$ and $\sigma$.  
A Rayleigh distribution with a similar
peak eccentricity is plotted with the dashed line.
\bigskip

\end{figure}

\subsection{Mass Distributions}

Naturally occurring protoplanet populations exhibit non-trivial distributions
in mass.  Before describing such a disk in the framework we have developed,
we clarify several points.

Protoplanets with different masses, or equivalently, different
radii, experience different viscous stirring rates. 
We decompose the total surface density in protoplanets, $\Sigma$, into
a differential quantity, $d\Sigma/dR$,
and write the excitation rate of a body with radius R as

\begin{equation}
\label{ratewvaryingmasses}
{\cal R}_x(e,R) \sim \int \frac{d\Sigma}{dR'} \frac{\Omega}{\rho R'} \frac{1}{\alpha^2}
\frac{e_H(R')}{e}~ dR'.
\end{equation}

\noindent
The identity $e_H(R') =(R'/R) e_{H}(R)$ when substituted into equation 
\ref{ratewvaryingmasses} yields

\begin{equation}
\label{varyingexciterate}
{\cal R}_x(e,R) \sim \frac{\Omega}{\rho R} \frac{1}{\alpha^2}
\frac{e_H(R)}{e} \int \frac{d\Sigma}{dR'}~ dR'.
\end{equation}

\noindent
In words, the excitation rate of one body only depends on the 
total surface density of all other bodies, regardless of the 
specific mass distribution.  
This differs from the assertion by \citet{GLS04} that only the 
most massive bodies contribute to the viscous stirring rate.
Equation \ref{varyingexciterate} seems to indicate
that there should be no distinction between big bodies and small bodies
since every body contributes to the viscous stirring.  A closer 
investigation uncovers the mass range of bodies that provide
significant stirring.

Eccentricity kicks of 
strength $e_k$ can occur at any combination of $M$ and $b$ that satisfies the
inverse square law of gravitation: $e_k \sim M(R') b(R')^{-2}$.  However,
the smallest impact parameter that contributes to a body's excitation 
is about $R_H$.
A minimum $b(R') \sim R_H$ sets a minimum mass for bodies 
to kick a body with mass $M$ by an amount $e_k$:

\begin{equation}
\label{mmin}
%e_k &\sim & \left( \frac{M_{\rm min}(R)}{M_{\odot}}\right)\left( \frac{a^2}{R_H^2}\right)
%\nonumber \\
M_{\rm min}(e_k,R)  \sim  \frac{e_k}{e_H} M
\end{equation}

\noindent
Likewise, a body can only be as far away as its radial position in the 
disk, $a$.  This sets a maximum mass, 

\begin{equation}
\label{mmax}
M_{\rm max}(e_k,R) \sim \left(\frac{e_k}{e_H} M \right)\frac{a^2}{R_H^2}.
\end{equation}

\noindent
For a choice of the most relevant kick strength, $e_k$, these
limits define the sizes of bodies that participate in the excitation of a body
with size $R$.

As a numerical confirmation of these results, 
we simulate a disk of planetesimals with a 
surface mass density $\sigma = 0.2 ~{\rm g~ cm^{-2}}$ and 120         
protoplanets.  In this case, we divide the protoplanets into 
two groups of different mass: sixty of mass 
$m_1 = 2\times 10^{24} ~{\rm g}$, and sixty of mass
$m_2 = 3.8 \times 10^{25} ~{\rm g}$.  These masses are within the 
limits set by equations \ref{mmin} \& \ref{mmax}.
We plot the absolute eccentricity distribution of each mass group 
binned separately in Figure 2.  Additionally, we plot the 
analytic distributions given
by $\sigma$, as specified above, and $\Sigma$, the sum of the surface
densities of both groups.  

It is clear that each group of protoplanets with the same mass
matches the analytic distribution well.
The offset between the peak of each group is due to the dependence of
the distribution on the Hill eccentricity of each body.
In general, the distribution for a swarm of protoplanets 
with a mass distribution is merely the sum of 
individual distributions for protoplanets of each mass.

%This generalization implies several results.  First, the surface density
%of protoplanets of all radii (subject to the limits of Eq. \ref{mmin} and \ref{mmax})
%contribute to the viscous stirring rate.
%Second, the eccentricity distribution 
%of bodies with radius $R$
%in units of their own Hill eccentricity is the same for all $R$.
%A numerical example illustrating these results is presented in
%Figure 2. This simulation includes planetesimals with a 
%surface mass density $\sigma = 0.2 ~{\rm g~ cm^{-2}}$ and one hundred 
%protoplanets (fifty with mass $m_1 = 2\times 10^{24} ~{\rm g}$, fifty with
%$m_2 = 3.8 \times 10^{25} ~{\rm g}$).

\begin{figure}

  \begin{center}

    \includegraphics[angle = -90, width = 0.7\columnwidth]{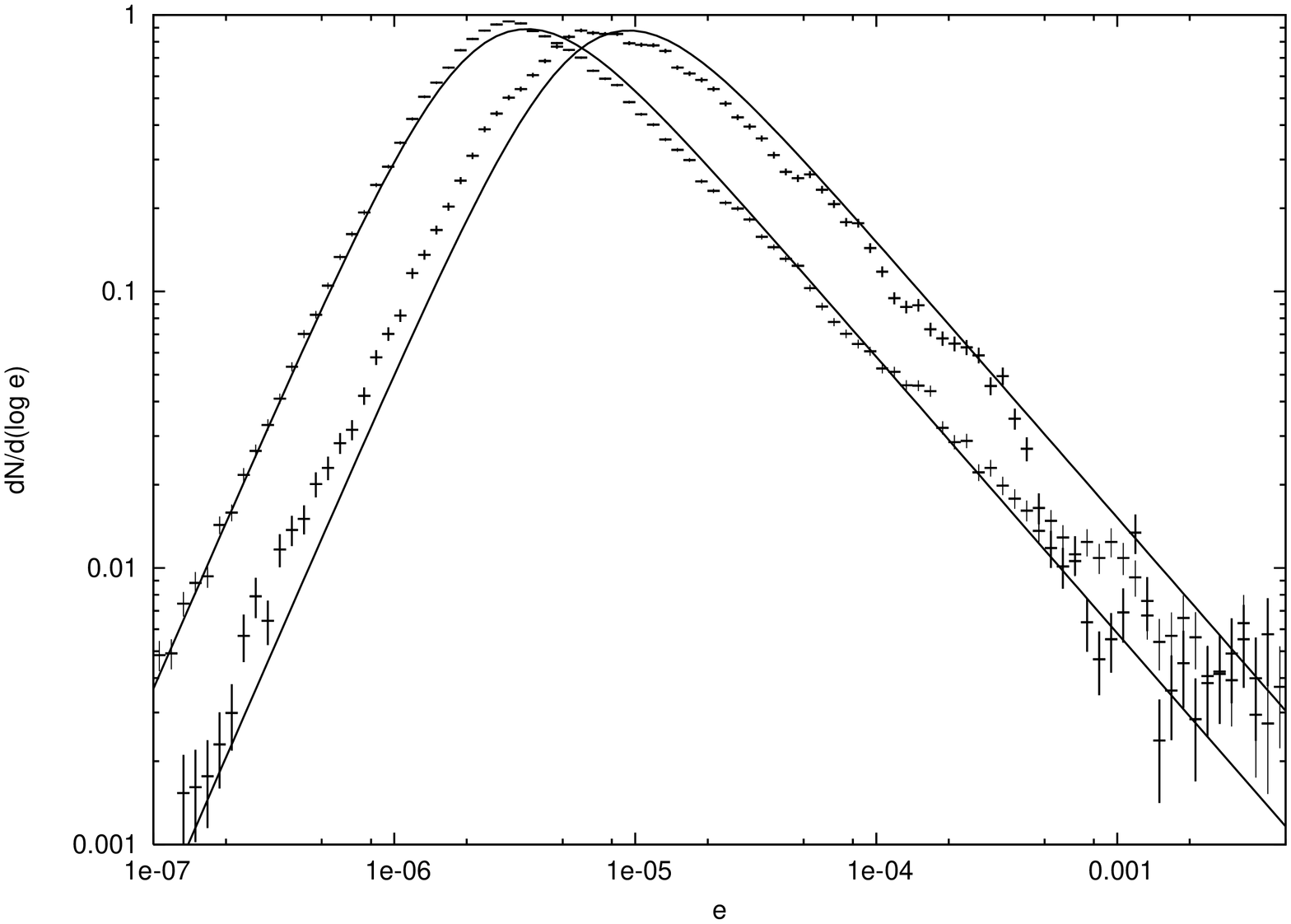}
    
  \end{center}
  
{\bf FIGURE 2:} A comparison of the results of a numerical simulation
of protoplanets in a perfectly bimodal mass distribution
($m_1 = 2\times 10^{24} ~{\rm g}, 
m_2 = 3.8 \times 10^{25} ~{\rm g}$).  We simulate
sixty bodies of each mass, for a total surface density in protoplanets
of $\Sigma \approx 0.003 ~{\rm g~ cm^{-2}}$ and 
a planetesimal surface density of $\sigma = 0.2 ~{\rm g~ cm^{-2}}$.
The eccentricities of each mass group are binned separately;
each distribution is a good match to  
equation \ref{thesolution} when scaled to
the appropriate Hill eccentricity.  The error bars are assigned 
following the same algorithm as Figure 1.
\bigskip
\end{figure}

\section{CONCLUSIONS}
\label{conclusions}

We presented an analytic model for the distribution function
of the eccentricities of a protoplanet population embedded in a 
shear-dominated planetesimal disk.  The eccentricity  distribution measured  
with numerical simulations matches the analytic result very well.

%We stress that we have not fit the model to the numerical
%measurement or vice versa.  
%The curves plotted in Figs.\ 1 and 2 have not been adjusted 
%in either direction.

Since we have manually inserted the dynamical friction rate
that we expect into the numerical simulations, this work does not test
our prescription of dynamical friction. 
However, the numerical and analytic representations of viscous 
stirring are completely independent.  Equation \ref{fulleq} uses
a viscous stirring rate involving only distant encounters.  In
our numerical simulations, Newton's laws dictate the 
protoplanet interactions directly without any simplifying assumptions.
The consistency of the two calculations proves that in a two-dimensional
shear-dominated disk, interactions between non-crossing orbits are
entirely responsible for setting the eccentricities of the protoplanets.
The analytic form of  the distribution function provides a simple way to test
similar hybrid numerical codes.

Several features of the distribution highlight interesting 
properties of the dynamics.
We reason in \S \ref{qualitative} that most protoplanets have
eccentricities $\sim (\Sigma/\sigma) e_H$, the value of $e$ where
the excitation and damping timescales are equal.  The distribution
function shows this to be true: 
the median and mean (up to a logarithmic factor) 
of any distribution are on the order of this equilibrium eccentricity.  
Higher moments of the distribution, however, are dominated by
the highest eccentricity bodies.  This signals that different 
statistics of the distribution can reflect different subsets of the overall population.
For example, the average ``thermal'' energy of the protoplanets
is represented by the root-mean-squared eccentricity,
$\langle e^2 \rangle$.  The fractionally fewer bodies with eccentricities 
close to $e_H$ dominate $\langle e^2 \rangle$ and thus, contain most of the 
energy.  

The shape of the distribution also merits discussion.
N-body integrations of a group of single
mass bodies show that their eccentricities
follow a Rayleigh distribution~\citep{IM92}.
For reference, we plot a Rayleigh distribution in Figure 1.
It is entirely inconsistent with our calculations.
This is not surprising.  In addition to simulating 
bodies in the regime of eccentricities that are large
compared to the Hill eccentricity, \citet{IM92}
do not include any effects that can balance the mutual 
excitations of their particles.
The dynamical friction in our simulations balances the viscous stirring 
and establishes the equilibrium distribution we derive.  

\acknowledgements
RS is an Alfred P. Sloan Fellow and a Packard Fellow.

\bibliographystyle{apj}
\bibliography{ms}

\appendix
\section{THE ANALYTIC DISTRIBUTION FUNCTION}
\label{proof}

Here we outline the evaluation of the equilibrium
equation, equation \ref{fulleq}, using the distribution 
function, equation \ref{thesolution}.  To simplify the notation, we rescale 
all eccentricities by $e_*$ and algebraically manipulate the coefficients
of each term in equation \ref{fulleq}.  We are left with the equivalent burden of 
proving that

\begin{equation}
g(e) = (1+e^2)^{-3/2}
\end{equation}

\noindent 
satisfies

\begin{equation}
2 \pi \frac{\partial g(e)}{\partial e} e + 4 \pi g(e)
= \int{\int{\frac{g(e)-g(e_n)}{|\vec{e}_n-\vec e|^{3}}~d^2 \vec{e_n}}}.
\end{equation}

The left-hand side is easy to compute:

\begin{equation}
\label{lhs}
{\rm L.H.S.} = \frac{4 \pi}{(1+e^2)^{3/2}}
-\frac{6\pi e^2}{(1+e^2)^{5/2}}.
\end{equation}

To integrate of the right-hand side, we translate the origin 
of the integration variables by $\vec e$ and rotate them
to align $\vec e$ with one of the coordinate axes.
In those coordinates:

\begin{equation}
\label{beforek}
I = \int_{-\infty}^{\infty} \int_{-\infty}^{\infty}
\frac{1}{(k^2+h^2)^{3/2}}
\left[ \frac{1}{(1+e^2)^{3/2}} - 
\frac{1}{(1+k^2+(h+e)^2)^{3/2}} \right]
 dk~ dh,
\end{equation}

\noindent
with $\vec e_n = \{k,h\}$.  

After the integration over $k$, we rewrite the integral
in terms of a new variable $h' \equiv (1+e^2)/(e h)$,

\begin{equation}
I = 
\int_{-\infty}^{\infty}
\left( \frac{2 e}{(1+e^2)^{5/2}}
+ \frac{8}{(1+e^2)^2} \frac{\partial^2 E(e^2 z /(1+e^2))}{\partial z^2}|h'| h'^2 \right) dh',
\end{equation}

\noindent
where, $z= -2h' - h'^2$, and $E(e^2 z /(1+e^2) )$ is 
the complete elliptic integral of the second kind.

We change the integration variable to $z$, taking care to evaluate the
integrand with the appropriate
branch of the double-valued relation $h'(z)$.
The integral evaluates to

\begin{equation}
I=\frac{4 \pi}{(1+e^2)^3}
+\frac{8}{(1+e^2)^2}\int_0^{1}\left[\frac{4-3z}{\sqrt{1-z}}\right]
\frac{\partial^2 E(e^2 z/(1+e^2))}{\partial z^2} dz.
\end{equation}

\noindent
With the second derivative
of the elliptic function expressed as a power series,
each term can be integrated over $z$.
The remaining power series in $e^2/(1+e^2)$ equals 

\begin{equation}
I=\frac{4 \pi}{(1+e^2)^3}
-\frac{3 \pi e^4}{(1+e^2)^4}
\left[ _2F_1\left(\frac{5}{2},1;3;\frac{e^2}{(1+e^2)}\right)
-\frac{1}{2}~ _2F_1\left(\frac{3}{2},2;3;\frac{e^2}{(1+e^2)}\right)
\right].
\end{equation}

\noindent
After additional algebraic manipulation, this result equals
the left hand side of the original equation (eq.\ \ref{lhs}).

\end{document}